\documentclass[twocolumn,amsmath,floatfix]{revtex4}
\usepackage{latexsym}
\usepackage{amssymb}
\usepackage{pdfpages}
\usepackage{graphics}
\begin{document}
\newcommand{\ja}{Jakuba\ss a-Amundsen }
\newcommand{\bfx}{\mbox{\boldmath $x$}}
\newcommand{\bfq}{\mbox{\boldmath $q$}}
\newcommand{\bfnabla}{\mbox{\boldmath $\nabla$}}
\newcommand{\bfalpha}{\mbox{\boldmath $\alpha$}}
\newcommand{\bfsigma}{\mbox{\boldmath $\sigma$}}
\newcommand{\bfeps}{\mbox{\boldmath $\epsilon$}}
\newcommand{\bfA}{\mbox{\boldmath $A$}}
\newcommand{\bfP}{\mbox{\boldmath $P$}}
\newcommand{\bfe}{\mbox{\boldmath $e$}}
\newcommand{\bfn}{\mbox{\boldmath $n$}}
\newcommand{\bfW}{{\mbox{\boldmath $W$}_{\!\!rad}}}
\newcommand{\bfM}{\mbox{\boldmath $M$}}
\newcommand{\bfI}{\mbox{\boldmath $I$}}
\newcommand{\bfJ}{\mbox{\boldmath $J$}}
\newcommand{\bfQ}{\mbox{\boldmath $Q$}}
\newcommand{\bfY}{\mbox{\boldmath $Y$}}
\newcommand{\bfp}{\mbox{\boldmath $p$}}
\newcommand{\bfk}{\mbox{\boldmath $k$}}
\newcommand{\bfks}{\mbox{{\scriptsize \boldmath $k$}}}
\newcommand{\bfqs}{\mbox{{\scriptsize \boldmath $q$}}}
\newcommand{\bfxs}{\mbox{{\scriptsize \boldmath $x$}}}
\newcommand{\bfalphas}{\mbox{{\scriptsize \boldmath $\alpha$}}}
\newcommand{\bfs}{\mbox{\boldmath $s$}_0}
\newcommand{\bfv}{\mbox{\boldmath $v$}}
\newcommand{\bfw}{\mbox{\boldmath $w$}}
\newcommand{\bfb}{\mbox{\boldmath $b$}}
\newcommand{\bfxi}{\mbox{\boldmath $\xi$}}
\newcommand{\bfzeta}{\mbox{\boldmath $\zeta$}}
\newcommand{\bfr}{\mbox{\boldmath $r$}}
\newcommand{\bfrs}{\mbox{{\scriptsize \boldmath $r$}}}
\newcommand{\bfps}{\mbox{{\scriptsize \boldmath $p$}}}

\renewcommand{\theequation}{\arabic{equation}}
\renewcommand{\thesection}{\arabic{section}}
\renewcommand{\thesubsection}{\arabic{section}.\arabic{subsection}}

\title{\Large\bf QED corrections to the parity-violating asymmetry\\ in high-energy electron-nucleus collisions}

\author{Xavier Roca-Maza$^{1,2,3,4}$ and D,~H.~Jakubassa-Amundsen$^5$}
\email{dj@math.lmu.de}
\email{xavier.roca.maza@fqa.ub.es}
\affiliation{$^1$Departament de F\'isica Qu\`antica i Astrof\'isica, Mart\'i i Franqu\'es, 1, 08028 Barcelona, Spain}
\affiliation{$^2$Dipartimento di Fisica ``Aldo Pontremoli'', Universit\`a degli Studi di Milano, 20133 Milano, Italy}
\affiliation{$^3$INFN, Sezione di Milano, 20133 Milano, Italy}
\affiliation{$^4$Institut de Ci\`encies del Cosmos, Universitat de Barcelona, Mart\'i i Franqu\'es, 1, 08028 Barcelona, Spain}
\affiliation{$^5$Mathematics Institute, University of Munich, Theresienstrasse 39, 80333 Munich, Germany}

\date{\today}

\vspace{1cm}

\begin{abstract}
The parity-violating asymmetry, including leading-order QED corrections to the Coulomb potential, is calculated non-perturbatively by solving the Dirac equation. At GeV collision energies and forward scattering angles, QED effects enhance the asymmetry by approximately 5\% for the recently measured nuclei $^{27}$Al, $^{48}$Ca, and $^{208}$Pb. The corrections result in a shift of the estimated neutron radius, leading to an increase in the inferred neutron skin thicknesses of these nuclei and, thus, to the pressure neutrons feel around nuclear saturation density.
\end{abstract}

\maketitle

{\it Introduction.--}Current knowledge of neutron distributions in atomic nuclei is limited by our capabilities to probe neutrons in a model-independent way. Parity violating elastic electron scattering by nuclei has been proposed and applied experimentally as a tool for a model independent determination of the weak charge distribution in ${}^{27}$Al\cite{An22}, ${}^{48}$Ca \cite{Ad22} and ${}^{208}$Pb \cite{Ab12,Ad21}. The main observable that is extracted from these experiments is the parity-violating asymmetry ($A_{\rm pv}$) which depends, apart from constants, on the momentum transfer, the weak and electric charge and the corresponding form factors (cf. Eq.~(1) in Ref.~\cite{Ad21}). Since the electromagnetic form factor is known for many nuclei from parity-conserving elastic electron scattering, this leaves the weak form factor the quantity that is actually determined in such experiments. According to the Standard Model, the weak charge of the neutron is close to $-1$ and that of the proton is close to $0$. Hence, this probe is particularly sensitive to neutrons and does not rely on modeling strongly interacting processes. The connection of the weak charge distribution with the neutron distribution in nuclei is straightforward and robust due to the sufficiently accurate knowledge on the weak form factors of the neutron and the proton (cf. Fig.~3 in \cite{Ad22}).

Due to the challenging nature of such experiments, the parity-violating asymmetry has been measured at only one or two momentum transfers for each nucleus. This limitation hinders a fully model-independent extraction of the weak-- and, consequently, neutron-- distribution in nuclei. However, these measurements are still considered sufficient to reasonably estimate the neutron root mean square radius, $\langle r_{\rm n}^2\rangle^{1/2}$, and define the neutron skin thickness, $\Delta r_{\rm np} \equiv \langle r_{\rm n}^2\rangle^{1/2} - \langle r_{\rm p}^2\rangle^{1/2}$. The neutron skin thickness is directly linked to the pressure experienced by neutrons at uniform densities near nuclear saturation, $\varrho_0 = 0.16$ fm$^{-3}$ \cite{Br00,Ce09,Ro11,Hu22}. This connection has significant implications for characterizing the QCD phase diagram at $T=0$ and baryon densities around $\varrho_0$, which, in turn, provide critical insights into the physics of neutron stars.

In addition to the relevance of this type of experiments in Nuclear Physics and Nuclear Astrophysics, parity-violating elastic electron scattering off nuclei can also be used as a test of the Standard Model \cite{Th13} through its sensitivity to the Weinberg mixing angle of the electroweak theory \cite{Ho12, Co22, Ca24} or by providing information on the weak charge radius in atomic nuclei for the analysis of atomic parity-violation data \cite{Po99}.

In earlier works analyzing the parity-violating asymmetry in ${}^{27}$Al\cite{An22}, ${}^{48}$Ca \cite{Ad22} and ${}^{208}$Pb \cite{Ab12,Ad21}, radiative corrections due to the $Z$ boson, in particular $\gamma-Z$ box contributions have been taken into account via a modification of the weak charge of the nucleus \cite{Go11,Er13,GH09}. However, within the electromagnetic sector, the elastic Coulomb scattering has never been corrected. In this work the vacuum polarization and the vertex plus self-energy QED corrections are included in order to assess their impact on the determination of the weak (neutron) radius of the nucleus keeping the weak charge value assumed in \cite{An22,Ad22,Ab12,Ad21}. This procedure allows to consistently account in the electromagnetic and weak sector for first-order corrections to the parity-violating asymmetry.


In addition, in the present work, the leading QED corrections to the Coulomb potential felt by the incident electrons are treated non-perturbatively and are absorbed into an exact calculation of the parity-violating spin asymmetry. This is necessary because Coulomb distortions are not negligible in experiments involving medium and heavy nuclei (cf. Fig.~1 in Ref.~\cite{Ho98}).

{\it Theory.--}The parity-violating asymmetry can be obtained from longitudinally polarized elastic electron scattering \cite{Ho01} as
\begin{equation}\label{1}
A_{\rm pv}\,=\, \frac{d\sigma/d\Omega(+)-d\sigma/d\Omega(-)}{d\sigma/d\Omega(+)+d\sigma/d\Omega(-)}
\end{equation}
where the differential scattering cross section $d\sigma/d\Omega(+)$ is for beam electrons with positive helicity  and $d\sigma/d\Omega(-)$ for those with negative helicity. Both differential cross sections are calculated by means of the  phase-shift analysis \cite{Lan}. The electron scattering states are obtained from solving the Dirac equation with the help of the Fortran code RADIAL \cite{Sal} (in atomic units, $\hbar=m=e=1$),
\begin{equation}\label{2}
[-ic\bfalpha \bfnabla + V_T(r) + V_{\rm vac}(r) + V_{\rm vs}(r) \pm A_{\rm w}(r)]\;\psi_\pm\,=\,E\,\psi_\pm,
\end{equation}
where $\psi_+ $ and $\psi_-$ denote the helicity states of the electron. The weak potential $A_{\rm w}$ is determined in terms of the weak density $\varrho_{\rm w} $ \cite{Ho98},

\begin{equation}\label{3}
A_{\rm w}(r)\,=\,\frac{G_Fc}{2\sqrt{2}}\;\varrho_{\rm w}(r),
\end{equation}
where $G_F$ is the Fermi coupling constant. The Coulomb target potential $V_T$ is derived from the nuclear ground-state charge distribution (see, e.g. Eq.~(3.4) in Ref.~\cite{Ho01}). $V_{\rm vac}$ is the Uehling potential as parametrized in \cite{FR76}. The potential $V_{\rm vs}$ for the radiative vertex plus self-energy (vs) correction is constructed from the relation between the first-order Born transition amplitude and the underlying potential \cite{Jaku24}, leading to
\begin{equation}\label{4}
V_{\rm vs}(r)\,=\,-\,\frac{2Z}{\pi}\int_0^\infty dq\;\frac{\sin(qr)}{qr}\;F_L(q)\;F_1^{\rm vs}(q),
\end{equation}
where $F_L$ is the nuclear ground-state charge form factor and $F_1^{\rm vs}$ is the electric form factor describing the vs correction in the first-order Born approximation \cite{Va00}.

The change of $A_{\rm pv}$ by the QED effects is quantified by means of the difference ($\Delta A_{\rm pv}$) and relative difference ($d A_{\rm pv}$) of the parity-violating spin asymmetries calculated, respectively, with ($A_{\rm pv(QED)}$) and without ($A_{\rm pv}$) the potentials $V_{\rm vac} +V_{\rm vs}$ in the Dirac equation (\ref{2}),
\begin{equation}\label{5}
dA_{\rm pv(QED)}\;\equiv\;\frac{A_{\rm pv(QED)}-A_{\rm pv}}{A_{\rm pv}}\equiv \frac{\Delta A_{\rm pv}}{A_{\rm pv}}.
\end{equation}
It should be emphasized that a consistent consideration of the QED effects on the spin asymmetry is not possible in Born approximation \cite{Jaku24}. The spin asymmetry is also not affected by soft bremsstrahlung \cite{Jo62,Jaku24}.

{\it The $^{208}$Pb nucleus.--}A Gaussian fit to experiment \cite{deV} was used for the nuclear charge distribution $\varrho_{\rm ch}$ which is well suited for high collision energies \cite{Ko21}. In fact, an analytic representation of $\varrho_{\rm ch}$ allows the calculation of the QED effects with much less numerical uncertainty.

For vacuum polarization, the asymptotic approximation of the Uehling potential by means of the first term in the large-$r$ expansion \cite{FR76} was used for $r\gtrsim 800$ fm. The vs potential was damped by an exponential tail \cite{Jaku24a} beyond 400 fm and was cut off near 900 fm in order to avoid an oscillatory behaviour of $V_{\rm vs}$ at very large distances (oscillations occur for an insufficient step number in the integration (\ref{4})).

\begin{figure}
\vspace{-1.5cm}
\includegraphics[width=11cm]{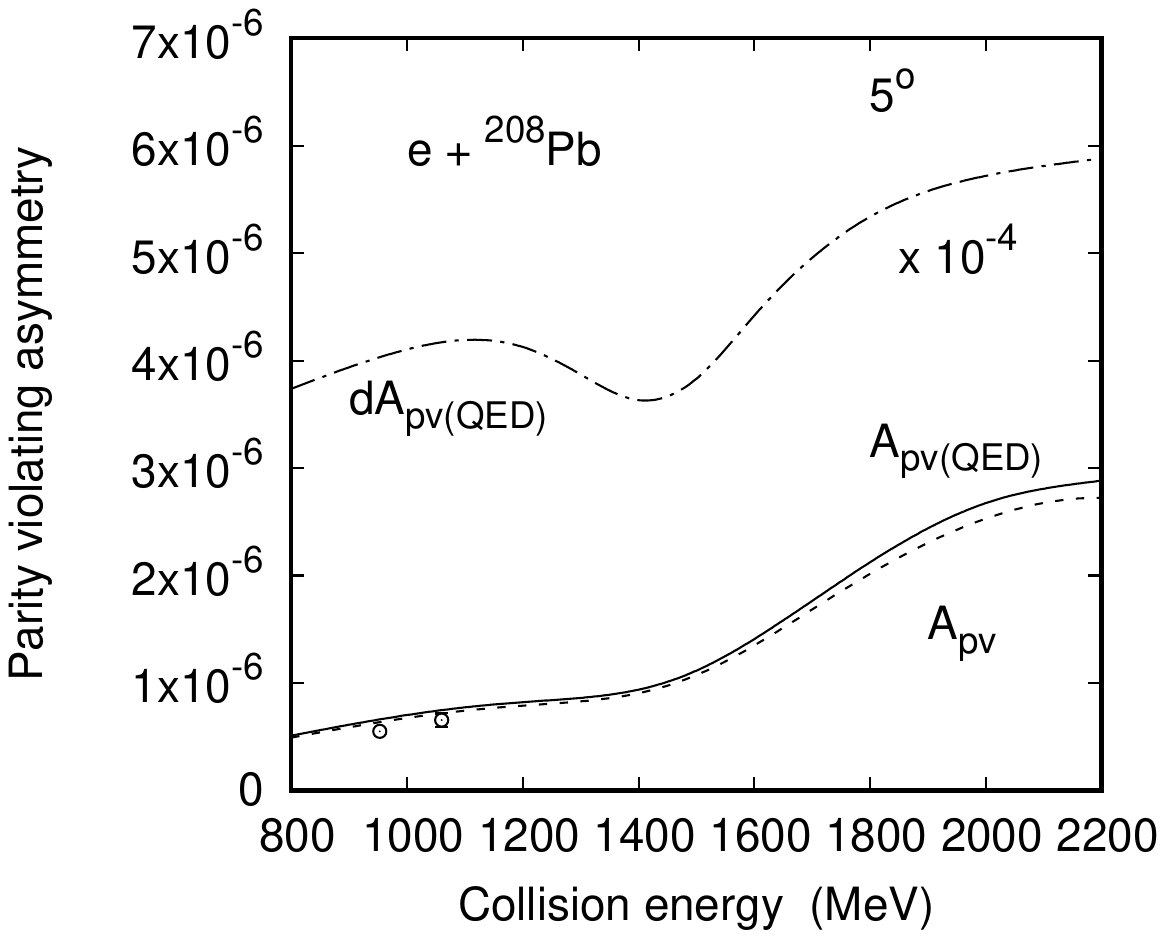}
\vspace{-1.cm}
\caption{Parity-violating spin asymmetry $A_{\rm pv}$ and its modification $dA_{\rm pv(QED)}$ by vacuum polarization and the vs correction for $e + ^{208}$Pb collisions at scattering angle $\vartheta_f =5^\circ$ as a function of collision energy. Shown is $A_{\rm pv} \;(----),\;A_{\rm pv(QED)}$ (---------), as well as $dA_{\rm pv(QED)}$, reduced by a factor of $10^{-4} \;(-\cdot -\cdot -)$. Included are the experimental data $(\odot)$ at 953 MeV \cite{Ad21} and at 1060 MeV \cite{Ab12}. \label{fig1}}
\end{figure}

The weak density $\varrho_{\rm w}$ was obtained from different nuclear Energy Density Functionals (EDFs) covering a representative range of realistic predictions on the nuclear densities \cite{Ba82}-\cite{Ro13}. In detail, neutron and proton point-like distributions were obtained and convoluted with the single neutron and proton weak form factors to obtain the total weak density of each nucleus \cite{Re21}.

Fig.~\ref{fig1} displays the parity violating spin asymmetry for $e$ + $^{208}$Pb collisions at a forward angle as function of the beam energy. It is seen that $A_{\rm pv}$ increases monotonously with energy, and the QED effects also increase basically  with energy. The structure in $dA_{\rm pv(QED)}$ mimics the diffraction structure in the elastic scattering cross section, and the QED correction amounts to about 4-6\%. It should be pointed out that $A_{\rm pv}$ is sensitive to the different nuclear models employed for the calculation of $\varrho_{\rm w}$. In the forward region where the experiments are carried out \cite{Ab12,Ad21}, the spread of theories is around 10\% (cf., e.g., the spread of the models in Fig.~3 of Ref.~\cite{Ad21}). However, the size of the QED correction $dA_{\rm pv(QED)}$ is quite insensitive to the underlying nuclear model, the sensitivity being about or below 1\% for the experimental kinematics considered here. Hence, the correction $dA_{\rm pv(QED)}$ is of the same order of magnitude as the model-dependence of $A_{\rm pv}$ and, thus, non-negligible. 

{\it The $^{48}$Ca nucleus.--}We have used a Gaussian as well as a Fourier-Bessel representation of the experimental ground-state charge distribution \cite{deV}. The weak potential has been derived theoretically using the same nuclear EDFs as employed for ${}^{208}$Pb. This ensures also for $^{48}$Ca  a reasonable variation of the weak charge distribution \cite{Ba82}-\cite{Ro13}.

\begin{figure}
\vspace{-1.5cm}
\includegraphics[width=11cm]{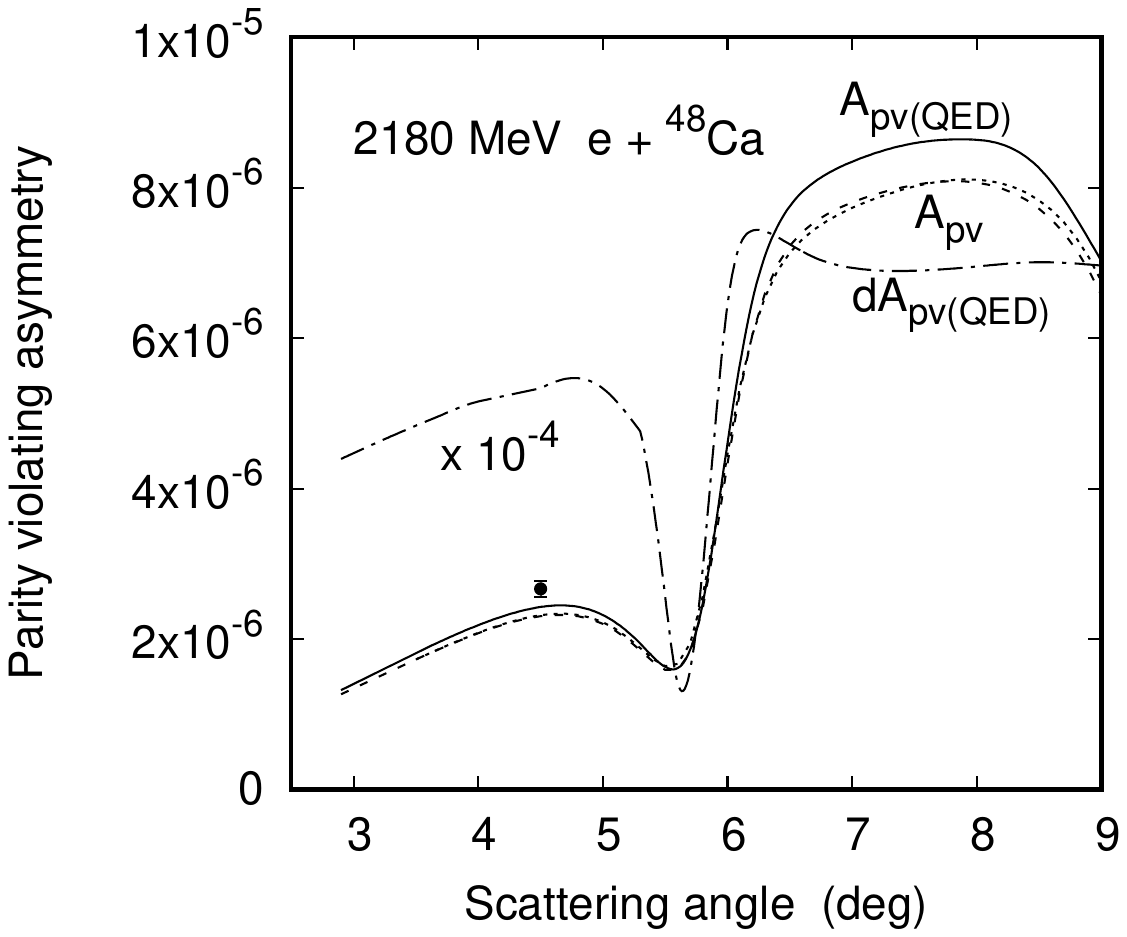}
\vspace{-1.0cm}
\caption
{Parity-violating spin asymmetry $A_{\rm pv}$ and its modification $dA_{\rm pv(QED)}$ for 2180 MeV $e + ^{48}$Ca collisions as a function of scattering angle $\vartheta_f$. Shown is $A_{\rm pv}$ resulting from a Gaussian charge distribution $(----),\; A_{\rm pv(QED)} $ (--------), as well as $dA_{\rm pv(QED)}$, reduced by the factor $10^{-4}\;(-\cdot -\cdot-)$. $A_{\rm pv}$ from the Fourier-Bessel representation of $\varrho_{\rm ch}$ is also shown $(\cdots\cdots)$. Included is the experimental datum point at $4.5^\circ$ ($\bullet$ \cite{Ad22}). \label{fig2}}
\end{figure}

The resulting parity-violating spin asymmetry is much alike for the two ground-state charge densities, deviations becoming apparent beyond $\vartheta_f=8^\circ$ for a collision energy of 2180 MeV. This is shown in Fig.~\ref{fig2} where the angular distribution of $A_{\rm pv}$ is displayed. Its modulation beyond $4.5^\circ$ by the diffraction structures is quite apparent. For the experimental point at $4.5^\circ$ \cite{Ad22}, the difference in $A_{\rm pv}$ between the two representations for $\varrho_{\rm ch}$ is 0.8\%.
The size of the QED correction $dA_{\rm pv(QED)}$ (using the Gaussian $\varrho_{\rm ch}$) is again about 5\% up to $5^\circ$, augmenting to 7\% at $9^\circ$.

\vspace{1cm}

{\it The $^{27}$Al nucleus.--}The decision to conduct measurements on the $^{27}$Al nucleus stemmed from the use of an aluminum alloy cell in the electron-proton scattering experiment \cite{An22}. This setup necessitated a study of the cell's influence on the parity-violating asymmetry of the proton and, consequently, on its weak charge as determined by the experiment.

In contrast to the previously discussed spin-zero nuclei, $^{27}$Al has nuclear spin $J=\frac{5}{2}$. For the spherical part of the nuclear ground-state charge distribution both a Fourier-Bessel parametrization and a two-parameter Fermi distribution were used \cite{deV}. For the experimental  beam energy, $E_{\rm i,kin}=1157 $ MeV, the two prescriptions for $\varrho_{\rm ch}$ lead to a deviation in $A_{\rm pv}$ of 0.3\% at the chosen scattering angle of $7.61^\circ$, increasing slightly with angle.

In order to estimate the importance of the higher multipoles, we have calculated the elastic scattering cross section within the distorted-wave Born approximation \cite{Jaku25} by employing the respective form factors provided in \cite{Ra21}. For the experimental geometry (1157 MeV and $7.61^\circ$) the contribution of the multipoles with $1\leq L\leq 5$ to the scattering cross section amounts to 1.1\%, where the quadrupole ($C2$) portion is by far the dominating one. This confirms the earlier result of Horowitz \cite{Ho12}, obtained at $E_{\rm i,kin} = 250$ MeV at an angle which corresponds to the experimental momentum transfer, that the contribution of  $C2$  to $A_{\rm pv}$ is negligible. However, when the angle is increased to $13^\circ$ (the position of the first diffraction minimum of $A_{\rm pv}$) the consideration of $C2$ increases the cross section by a factor of two, while the magnetic contribution is still small ($\sim 1\%)$. This explains the large change of $A_{\rm pv}$ found in \cite{Ho12} near the minimum upon adding $C2$. The $\varrho_{\rm w}$ was obtained with the EDFs of Refs.~\cite{Ba82}-\cite{Ro13} imposing the blocking approximation for the unpaired proton. This approximation is reasonable to model the weak charge distribution for the different models and to calculate $dA_{\rm pv (QED)}$.     

\begin{table}[!h]
\begin{center}
  \caption{Theoretical $A_{\rm pv}$ and $\Delta A_{\rm pv}\equiv A_{\rm pv(QED)}-A_{\rm pv}$ for the targets $^{27}$Al, $^{48}$Ca and $^{208}$Pb are shown in this table. Ranges are given for $A_{\rm pv}$ indicating the magnitude of the model dependence on this observable. The value within parenthesis in $\Delta A_{\rm pv}$ is an estimation of the theoretical uncertainty of the QED corrections. Experimental data and corresponding references are given in the last two columns.}
\label{tab1}
\begin{tabular}{rrccccc}
\hline\hline
Target & $E_{\rm i,kin}$  & $\vartheta_f$ & $A_{\rm pv}$&$\Delta A_{\rm pv}$& $A_{\rm pv}({\rm exp})$&Ref.\\
&(MeV)&(deg.)&(ppm)&(ppm)&(ppm)&\\ \hline
$^{27}$Al&1157& $7.61$&$[1.95,2.05]$  &$0.103(2)$&$2.16(19)$& \cite{An22}\\
$^{48}$Ca& 2180&$4.51$&$[2.06,2.40]$  &$0.120(10)$&$2.67(11)$& \cite{Ad22}\\
$^{208}$Pb& 953&$4.7 $&$[0.556,0.591]$&$0.022(1)$&$0.550(18)$&\cite{Ad21}\\
         &1060&$5   $&$[0.671,0.729]$&$0.029(1)$&$0.656(62)$&\cite{Ab12}\\
 \hline\hline
\end{tabular}
\end{center}
\end{table}

{\it Comparison with experiment.--}The values of $A_{\rm pv}$ and the corresponding QED-induced shift,  $\Delta A_{\rm pv}$, predicted by the models utilized in this study and calculated for the experimental geometries, are summarized in Table \ref{tab1}. The range predicted by the models for $A_{\rm pv}$ shown within square brackets is in agreement with previous calculations (see, e.g., Refs.\cite{Ba12, Ad21, Re22}) and none of them overlaps the experimentally determined central values. The correction $d A_{\rm pv(QED)}$, about 5\% in all cases, is almost model-independent as can be seen from the estimated theoretical uncertainties of $\Delta A_{\rm pv}$ (given within parenthesis). In detail, taking the calculated central values for $A_{\rm pv}$ as are used in the figures, the QED effects change $A_{\rm pv}$ to $2.15\times 10^{-6}$  in ${}^{27}$Al, $2.43\times 10^{-6}$  in ${}^{48}$Ca and $6.07\times 10^{-7}$  in ${}^{208}$Pb (most recent experiment) and hence are not negligible. When an average over the experimental resolution is made, using the appropriate acceptance functions \cite{Ad21,Ad22}, the theoretical central values for $A_{\rm pv(QED)}$ are modified to $2.36\times 10^{-6}$ for $^{48}$Ca and $5.98 \times 10^{-7}$ for $^{208}$Pb at 953 MeV. The systematic effect of $\Delta A_{\rm pv}$ increasing the theoretical predicted value of $A_{\rm pv}$ improve the theoretical description of $^{27}$Al and $^{48}$Ca. For the case of $^{208}$Pb, the QED corrections push further the theoretical prediction with respect to the experimental central values. Thus, understanding the experimental results on $^{208}$Pb remains a challenge in the context of the current approximation. 

The corrections obtained here together with our current theoretical understanding of the neutron skin imply that the latter must be larger than extracted from previous analysis and, consequently, the pressure of neutron matter around saturation density must also be larger \cite{Br00}-\cite{Ro11}. For the determination of the neutron skin thickness $\Delta r_{\rm np}$ the correlations between $A_{\rm pv}$ and $\Delta r_{\rm np}$ for each target, as provided in the literature \cite{Ad22,Ad21,Re22}, are used. This method is reliable since our QED changes are basically model-independent. Specifically, for ${}^{27}$Al a $\Delta r_{\rm np}=-0.04(12)$ fm \cite{An22} was previously reported and, according to Eq.~(6) in Ref.~\cite{An22}, this value must be instead $\Delta r_{\rm np}=0.02(12)$ fm if QED corrections were taken into account. For the case of ${}^{48}$Ca a $\Delta r_{\rm np}=0.121(35)$ fm was reported in \cite{Ad22} and, according to the results of Ref.~\cite{Re22}, the value must now be increased to $\Delta r_{\rm np}=0.161(35)$ fm. Finally, for the case of ${}^{208}$Pb, according to Fig.~3 of Ref.~\cite{Ad21}, the neutron skin must be shifted by 0.09 fm with respect to the reported value. That is, $\Delta r_{\rm np}=0.38(7)$ fm. The MREX experiment at Mainz will measure  $A_{\rm pv}$ in ${}^{208}$Pb with an expected accuracy of about $\pm 0.03$ fm on its neutron radius \cite{Th19}. It is also foreseen to measure ${}^{48}$Ca. The results of such experiments will be crucial to confirm our approximation here.

{\it Conclusion.--}We have provided a non-perturbative calculation of the QED effects, together with an exact treatment of Coulomb distortion. For collision energies ranging between 950 and 2200 MeV and scattering angles between $3^\circ$ and $9^\circ$ an increase of the parity-violating spin asymmetry by some 5\% was found, irrespective of the target species and of the model employed. For ${}^{27}$Al and ${}^{48}$Ca, this brings $A_{\rm pv}$ closer to experiment, while it is vice versa for $^{208}$Pb. As a consequence, the present work suggests that the current nuclear structure theory is well suited for the description of the neutron skins in ${}^{27}$Al and ${}^{48}$Ca while it may have some difficulties in accomodating a large skin such as the one estimated here for ${}^{208}$Pb.  

Apart from these QED effects, the $\gamma-Z$ box diagram could contribute to the weak radius and not only to the weak charge as assumed so far in our calculations. As a consequence, it would further correct the value of $A_{\rm pv}$. We have validated that, when making restriction to transient nuclear excitations of angular momentum $L\leq 3$ and excitation energies up to 30 MeV, these dispersive changes to the parity-violating spin asymmetry are much smaller than the QED corrections and hence can be neglected. 

{\it Acknowledgments.--}XRM acknowledges support by MICIU/AEI/10.13039/501100011033 and by FEDER UE through grants PID2023-147112NB-C22; and through the ``Unit of Excellence Mar\'ia de Maeztu 2025-2028" award to the Institute of Cosmos Sciences, grant CEX2024-001451-M. Additional support is provided by the Generalitat de Catalunya (AGAUR) through grant 2021SGR01095.

\vspace{1cm}


\begin{thebibliography}{99}

\bibitem{An22} D.Androi\'{c} et al. (Qweak Collaboration), Phys. Rev. Lett. {\bf 128}, 132501 (2022).

\bibitem{Ad22} D.Adhikari et al. (CREX Collaboration), Phys. Rev. Lett. {\bf 129}, 042501 (2022).

\bibitem{Ab12} S. Abrahamyan et al. (PREX Collaboration), Phys. Rev. Lett. {\bf 108}, 112502 (2012).

\bibitem{Ad21} D.Adhikari et al. (PREX Collaboration), Phys. Rev. Lett. {\bf 126}, 172502 (2021).


\bibitem{Br00} B.~A. Brown, Phys. Rev. Lett. {\bf 85} (2000) 5296-5299.
  
\bibitem{Ce09} M. Centelles, X. Roca-Maza, X. Viñas, M. Warda. Phys. Rev. Lett. {\bf 102} 122502 (2009). 

\bibitem{Ro11} X. Roca-Maza, M. Centelles, X. Viñas, M. Warda, Phys.Rev.Lett. {\bf 106} 252501 (2011).

\bibitem{Hu22} Baishan Hu, et al. Nature Physics {\bf 18}, 1196–1200 (2022).
  
\bibitem{Th13} M.Thomson, {\it Modern Particle Physics} (Cambridge University Press, Cambridge, 2013).
  
\bibitem{Ho12} C.J.Horowitz et al., Phys. Rev. C {\bf 85}, 032501(R) (2012).
  
\bibitem{Ca24} Matteo Cadeddu, Nicola Cargioli, Jens Erler, Mikhail Gorchtein, Jorge Piekarewicz, Xavier Roca-Maza, Hubert Spiesberger, Phys. Rev. C {\bf 110} 035501 (2024).

\bibitem{Co22} M.Alzori Corona, M.Cadeddu, N.Cargioli, P.Finelli and M.Vorabbi, Phys. Rev. C {\bf 105}, 055503 (2022).
  
\bibitem{Po99} S. J. Pollock, M. C. Welliver, Physics Letters B {\bf 464} 177-182 (1999)
  
\bibitem{Go11} M. Gorchtein, C. J. Horowitz, and M. J. Ramsey-Musolf, Phys. Rev. C 84, 015502 (2011).
  
\bibitem{Er13} J. Erler and S. Su, Prog. Part. Nucl. Phys. 71, 119 (2013).
  
\bibitem{GH09} M.Gorchtein and C.J.Horowitz, Phys. Rev. Let. {\bf 102}, 091806 (2009).

\bibitem{Ho98} C.J.Horowitz, Phys. Rev. C {\bf 57}, 3430 (1998).
  
\bibitem{Ho01} C. J. Horowitz, S. J. Pollock, P. A. Souder, R. Michaels, Phys. Rev. C {\bf 63}, 025501 (2001). 
  
\bibitem{Lan} V.B.Berestetskii, E.M.Lifshitz and L.P.Pitaevskii, {\it Quantum Electrodynamics} (Course of Theoretical Physics vol.4) 2nd edition (Elsevier, Oxford, 1982).

\bibitem{Sal} F.Salvat, J.M.Fern\'{a}ndez-Varea and  W.Williamson Jr.,  Comput. Phys. Commun.  {\bf 90}, 151 (1995).


\bibitem{FR76} L.W.Fullerton and G.A.Rinker Jr., Phys. Rev. A {\bf 13}, 1283 (1976).

\bibitem{Jaku24} D.H.Jakubassa-Amundsen, J. Phys. G {\bf 51}, 035105 (2024).

\bibitem{Va00} M.Vanderhaeghen, J.M.Friedrich, D.Lhuillier, D.Marchand, L.Van Hoorebeke, J.Van de Wiele, Phys. Rev. C {\bf 62}, 025501 (2000).

\bibitem{Jo62} W.R.Johnson, C.O.Carroll and C.J.Mullin, Phys. Rev. {\bf 126}, 352 (1962).

\bibitem{deV} H.De Vries, C.W.De Jager and  C.De Vries,  At. Data Nucl. Data Tables  {\bf 36}, 495 (1987). 

\bibitem{Ko21} O.Koshchii, M.Gorchtein, X.Roca-Maza and H.Spies\-berger, Phys. Rev. C {\bf 103}, 064316 (2021).

\bibitem{Jaku24a} D.H.Jakubassa-Amundsen, arXiv:2404.03445 [nucl-th].

\bibitem{Ba82} J. Bartel, P. Quentin, M. Brack, C. Guet, H.-B. Hakansson, Nucl. Phys. A {\bf 386} 79 (1982).  
  
\bibitem{Ro12} X. Roca-Maza, G. Col\`o, H. Sagawa, Phys.Rev. C {\bf 86} 031306 (2012).

\bibitem{Ro13} X. Roca-Maza, et al. Phys. Rev. C {\bf 87}, 034301 (2013)
  
\bibitem{Re21} P.-G.Reinhard, X.Roca-Maza and W.Nazarewicz, Phys. Rev. Lett. {\bf 127}, 232501 (2021).  



\bibitem{Jaku25} D.H.Jakubassa-Amundsen, Eur. Phys. J. A 61, 73 (2025).

\bibitem{Ra21} R.A.Radhi, A.A.Alzubadi and  N.S.Manie, Nucl. Phys. A {\bf 1015}, 122302 (2021).

\bibitem{Re22} P.-G.Reinhard, X.Roca-Maza and W.Nazarewicz, Phys. Rev. Lett. {\bf 129}, 232501 (2022).

\bibitem{Ba12} S.Ban, C.J.Horowitz and R.Michaels, J. Phys. G {\bf 39}, 015104 (2012).

\bibitem{Th19} M. Thiel, C. Sfienti, J. Piekarewicz, C. J. Horowitz and M. Vanderhaeghen, J. Phys. G: Nucl. Part. Phys. {\bf 46} 093003 (2019). 

\end{thebibliography}
\end{document}